# DE-LEACH: Distance and Energy Aware LEACH

Surender Kumar
University of Petroleum and
Energy Studies, India

M.Prateek, N.J.Ahuja
University of Petroleum and
Energy Studies, India

Bharat Bhushan
Guru Nanak Khalsa College,
Yamunanagar, India

## ABSTRACT
Wireless sensor network consists of large number of tiny sensor nodes which are usually deployed in a harsh environment. Self configuration and infrastructure less are the two fundamental properties of sensor networks. Sensor nodes are highly energy constrained devices because they are battery operated devices and due to harsh environment deployment it is impossible to change or recharge their battery. Energy conservation and prolonging the network life are two major challenges in a sensor network. Communication consumes the large portion of WSN energy. Several protocols have been proposed to realize power-efficient communication in a wireless sensor network. Cluster based routing protocols are best known for increasing energy efficiency, stability and network lifetime of WSNs. Low Energy Adaptive Clustering Hierarchy (LEACH) is an important protocol in this class. One of the disadvantages of LEACH is that it does not consider the nodes energy and distance for the election of cluster head. This paper proposes a new energy efficient clustering protocol DE-LEACH for homogeneous wireless sensor network which is an extension of LEACH. DE-LEACH elects cluster head on the basis of distance and residual energy of the nodes. Proposed protocol increases the network life, stability and throughput of sensor network and simulations result shows that DE-LEACH is better than LEACH.

## 1. INTRODUCTION
With the recent advances in electromechanical devices and wireless communication technology low cost and low power consumption tiny sensors are available. Sensors collect the data from its surrounding area, carry out simple computations, and communicate with other sensors or with the base station (BS) [1]. Main challenges of wireless sensor network are energy efficiency, stability and network lifetime because the battery capacity of the sensor nodes is limited and due to harsh deployment of the sensor nodes, it is impractical to change or recharge the battery. Clustering of the network is an alternative to increase the energy efficiency and performance of the network. In hierarchical architecture sensing region is divided in to a number of clusters and a set of nodes are periodically elected as cluster head (CH). Cluster head collects the data from non-CH nodes, aggregate it and then send it to the sink for further processing [2] [3][4] .Clustering thus evenly distribute the energy load, reduce the energy consumption and increase the network life [4], [5], [6], [7], [8], [9], [10], [11], [12]. In this paper a novel energy efficient cluster based scheme is proposed for homogeneous network. The scheme suggests a new cluster head (CH) election mechanism based on the residual energy and the average distance of the nodes from the sink. This protocol is an improved version of LEACH protocol presented in [5] and simulation result shows that the proposed scheme is better than LEACH. The rest of this paper is divided into sections as follows. Section 2 explains the related works. Section 3 describes the wireless sensor network model, Section 4 describes radio energy model, Section 5 explains DE-LEACH protocol, simulation results are explored in Section 6 and finally the paper is concluded in Section 7.

## 2. RELETED WORK
Routing in a wireless sensor network is more challenging than the traditional networks due to sensors smaller memory, less processing power and constrained energy supply. In the past few years many new routing protocols have been devised for wireless sensor network [2], [3], [4], [7]. LEACH [5] is probably the first and most important clustering protocol for wireless sensor network which out performs the classical routing protocol by using adaptive clustering scheme. In this sensor nodes communicate with each other by using a single-hop communication. It operates in rounds and each round is further sub divided into two phases: setup phase and steady-state phase. In the setup phase each node generates a random number between 0 and 1 and if this random number is less than a particular threshold value T (n) which is given by Equation (1) then the node becomes a cluster head for the current round. After the CH election, non CH nodes select a CH which is nearest to them and then CH creates a TDMA schedule for its members so that they can transmit their data to the CH. Cluster head further aggregates the data received from its members and after that transmit it to base station. Author of [5] through simulation shows that only 5 percent of the nodes need to become a cluster head in a round. One of the disadvantages of the LEACH is that the cluster head rotations do not take into account the remaining energy of sensor nodes. LEACH-centralized (LEACH-C) is an enhancement over LEACH protocol [5]. LEACH-C uses a centralized algorithm, to distribute clusters throughout the sensing region. Set up phase of LEACH-C is different than LEACH. In setup phase of LEACH-C, nodes send their location and energy information by using a GPS to base station. After getting this information base station calculates the average energy of the sensor network. Nodes which have energy below this average energy cannot become the cluster head for the current round. From the remaining nodes base station finds out the best candidate for the cluster head election.

$$T(n) = \begin{cases} \dfrac{p}{1- p \times \left( r \ mod\dfrac{1}{p}\right)} & \text{If } n \, \varepsilon \, G \\ 0 & \text{otherwise} \end{cases} \qquad (1)$$

Here G denotes the set of nodes that are not selected as a cluster head in last $\frac{1}{p}$ rounds and r is the current round. One of the disadvantages of LEACH is that cluster heads rotations do not take into account the remaining energy of sensor nodes. A node may not have sufficient energy to complete a round and may be selected as a CH.

E-LEACH [6] applied both LEACH and a new approach for cluster head selection. When the remaining energy of a node is larger than 50% of the initial energy then LEACH algorithm is used as in Equation (1). Otherwise a new approach which considers the remaining energy of each node is applied for cluster head selection as shown in Equation (2).

$$T(n) = \begin{pmatrix} \frac{p}{1-p\times(r\ mod\ \frac{1}{p})} \times \left(2p \times \frac{E_{residual}}{E_{init}}\right) & \text{If n } \varepsilon \text{ G} \\ 0 & \text{otherwise} \end{pmatrix} \quad (2)$$

Here p is the percentage of nodes that can become cluster head, $E_{residual}$ is remaining energy and $E_{init}$ is the initial energy of a node.

In PEGASIS [8] instead of forwarding the packets from many cluster heads as like in LEACH, each node will form a chain structure to the base station through which the data is forwarded to the base station. PEGASIS achieve energy efficiency by transmit data to only one of its neighbor node. Here the collected data is fused and this will be further forwarded to its immediate one hop neighbor. Since all the nodes are doing the data fusion at its place there is no rapid power depletion of nodes which are present near to base station

HEED [9] is another popular energy efficient clustering algorithm which periodically elects cluster heads based on the hybrid of two parameters: residual energy of sensor nodes and intra cluster communication cost as a function of neighbor proximity or cluster density.

SEP [13] is a stable election protocol which study the impact of heterogeneity in a sensor work. There are two types of nodes in the network: normal and advanced node. Advanced nodes create a source of heterogeneity in the network and have more energy than normal nodes. Weighted election probabilities of the nodes are used for the election of cluster head according to the remaining energy of the nodes. Threshold for normal and advanced nodes are calculated by using weighted probabilities.

DEEC [14] is an energy-aware adaptive clustering protocol used in heterogeneous wireless sensor networks. In DEEC, every sensor node independently elects itself as a cluster-head based on its initial and residual energy. To control the energy expenditure of nodes by means of an adaptive approach, DEEC use the average energy of the network as the reference energy. Thus, DEEC does not require any global knowledge of energy. Unlike SEP and LEACH, DEEC can perform well in multi-level heterogeneous wireless sensor networks. DBCP [15] is a protocol for three level heterogeneous networks which uses a new cluster head election scheme based on the initial energy and average distance of the nodes from the sink.

TEEN [10] is a cluster based protocol for time critical applications. Cluster heads in TEEN broadcast two values known as hard and soft threshold. Hard threshold value is for sensed attributes and soft threshold is for small changes in the sensed value. Sensor nodes do the continuous sensing but they transmit the data less frequently. When the sensed value of nodes is more than hard threshold then they switch on their transmitter. This value is stored in an intermediate variable and in the current round nodes will transmit the data again only when both of the following conditions are satisfied.

- Sensed attribute current value is more than the hard threshold.
- Sensed attribute differs from intermediate variable by a value which is greater than or equal to soft threshold.

Hard threshold reduces the number of transmissions to transmit only when the sensed value is in the region of interest and soft threshold further reduces the transmission when there is a little or no change in the sensed value.

APTEEN is an extension to TEEN protocol where periodicity and threshold value is changed according to user needs and application type [11]. APTEEN uses three different query types namely: historical, one-time and persistent. Historical query analyses the past data, one-time query is for taking a network snapshot view and persistent query is for monitoring an event to a period of time. APTEEN is equally good for reactive and proactive policies. Additional complexity of count time and thresholds is the disadvantage of this scheme.

## 3. WIRELESS NETWORK MODEL
This section describes the wireless sensor network model which has been used in this paper. Model has n sensor nodes which are randomly deployed in a 100 x 100 square meters region as shown in Figure 1. Various assumptions make about the network model and sensors are as follows.

- Sensors are deployed randomly in the region.
- Base station and sensors become stationary after deployment and base station is 75 meters away from the sensing region.
- Sensors are location unaware i.e. they do not have any information about their location.
- Sensors continuously sense the region and they always have the data for sending to base station.
- Battery of the sensors cannot be changed or recharged as the nodes are densely deployed in a harsh environment.
- All the sensors have same amount of energy and processing capabilities i.e. network is homogeneous.

In the cluster based protocol cluster head does the communication with its members and also the aggregation of collected data to save the energy. If n sensors are distributed uniformly in M x M region and k is the optimal number of clusters per round. Then on average $\frac{n}{k}$ nodes are per cluster (one cluster head and $(\frac{n}{k} - 1)$ non-cluster head nodes).As base station is located far from the sensing region and according to [5] the energy dissipated in the cluster head during a single round follows the multipath path model ($d^4$ power loss) which is given by Equation (2).

$$E_{CH} = \left(\frac{n}{k} - 1\right) L. E_{elec} + \frac{n}{k} L. E_{DA} + L. E_{elec} + L. \varepsilon_{mp} d_{BS}^4 \quad (2)$$

Where L is the no of bits of the data message, $d_{BS}$ is the distance between base station and cluster head, $E_{DA}$ is the energy required to perform data aggregation in a round and k is the number of clusters. Since cluster members transmit data to only its cluster head therefore energy dissipated in a non cluster head follows the free space path model ($d^2$ power loss) and it is given by Equation (3).

$$E_{NCH} = L. E_{elec} + L. \epsilon_{fs}\, d_{CH}^2 \quad (3)$$

Here $d_{CH}$ is the distance between nodes and cluster head. The energy depleted in a cluster is given by Equation (4) and total energy dissipated in the network will be given by Equation (5).

$$E_{cluster} = E_{CH} + \left(\frac{n}{k} - 1\right) E_{NCH} \approx E_{CH} + \frac{n}{k} E_{NCH} \quad (4)$$

$$E_{total} = L(2nE_{elec} + nE_{DA} + k \epsilon_{mp}\, d_{BS}^4 + n \epsilon_{fs}\, d_{CH}^2) \quad (5)$$

The optimal number of clusters can be found by finding the derivative of $E_{total}$ with respect to k and equating it to zero.

$$k_{opt} = \frac{\sqrt{n}}{\sqrt{2\pi}} \sqrt{\frac{\epsilon_{fs}}{\epsilon_{mp}}} \frac{M}{d_{BS}^2} \quad (6)$$

$p_{opt}$ is the optimal probability of a node to become cluster head and it can be calculated as follows:

$$p_{opt} = \frac{k_{opt}}{n} \quad (7)$$

## 4. RADIO ENERGY MODEL

This paper has used the radio dissipation energy model of [5]. Radio model consists of three parts: transmitter, power amplifier and the receiver. Two propagation models: free space ($d^2$ power loss) and multipath fading ($d^4$ power loss) channel are used. The energy spent for transmitting an L bit data message to a distance d is given by Equation (8)

$$E_{TX(L,d)} = \begin{pmatrix} L \times E_{elec} + \epsilon_{fs} \times d^2 & if\ d < d_0 \\ L \times E_{elec} + \epsilon_{mp} \times d^4 & if\ d \geq d_0 \end{pmatrix} \quad (8)$$

$E_{elec}$ is the electricity spent to run the transmitter or receiver circuitry. The parameters $\varepsilon_{mp}$ and $\varepsilon_{fs}$ denotes the amount of energy spent per bit in the radio frequency amplifier according to the cross over distance $d_0$ which is given by Equation (9).

$$d_0 = \sqrt{\frac{\epsilon_{fs}}{\epsilon_{mp}}} \quad (9)$$

The energy expanded to receive an L bit message is given by Equation (10).

$$E_{RX(L,d)} = L \times E_{elec} \quad (10)$$

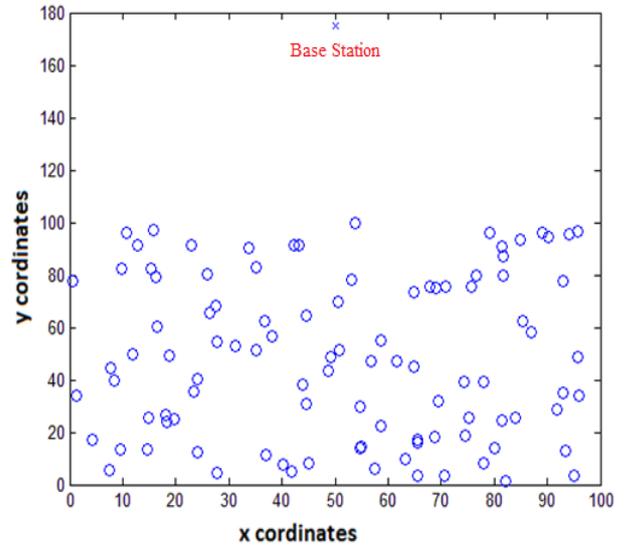

**Figure 1 Sensing Region of DE-LEACH**

**(o – Normal Node, x –Base Station)**

## 5. DE-LEACH

DE-LEACH (Distance & Energy Aware LEACH) elects cluster head on the basis of initial energy, residual energy and distance of the nodes from base station. Cluster heads further act as routers to base station and aggregation of the collected data of its members are also performed by them. DE-LEACH operation is divided into rounds and each round further consists of two phases: setup and steady-state. In setup phase DE-LEACH does the cluster heads election, cluster formation and determine the communication schedules for its members. When cluster formation is completed in setup phase DE-LEACH switches to steady state phase. In this phase it completes the data sensing, data transmission to cluster heads and aggregation of the collected data of its members.

The minimum amplifier energy of a transmitter or receiver is directly proportional to the square of the distance [5]. Thus long distance transmission increases the energy consumption in a WSN. Communication energy of WSNs is much more than sensing and computation energy. For example energy consumed in transmitting one bit data to a receiver at 100 meter away is equal to execute 3000 instructions [2]. The ratio of energy consumed in communication of one bit to process the same bit is in the range of 1,000 – 10,000 [7]. Thus in order to save energy and prolonging the network lifetime, network traffic and transmission distance should be decreased [5].There is a considerable difference in the energy of the far and near nodes of sensor network when it operates for some time. [15]. Authors of LEACH through simulation shows that in a round if the percentages of cluster heads are between 3 and 5 then they are more energy efficient [5]. To save energy of sensor network DE-LEACH uses different schemes for electing cluster heads from nearer and far nodes. On the basis of average distance from the base station DE-LEACH divides the entire sensing region in two parts. The first region whose distance is less than or equal to average distance from base station uses a cluster head election scheme which is based on the node distance from the base station and the percentage of nodes that can become cluster head is slightly more than 6 percentages. Similarly the second region whose distance is greater than the average distance from the base station use a scheme based on the residual and initial energy of the nodes and the percentage of nodes that can become cluster head is slightly more than 3 percentages. The reason behind using two different schemes for the cluster head election is that it will increase the chances of nodes to become cluster head in a round which are in the mid of sensing region than the other nodes. Initially after deployment each sensor node sends a message to the base station. Based on this signal strength the distance of the node from the base station can be calculated.

If $d_i$ is the distance between a node and the base station then the average distance ($d_{avg}$) of nodes from the sink [15] can be found by using the Equation 10.

$$d_{avg} = \frac{1}{n}\sum_{i}^{n} d_i \qquad (10)$$

The Value of $d_{avg}$ can be further approximated as [15]

$$d_{avg} \cong d_{CH} + d_{BS} \qquad (11)$$

If the distance $d_i$ of a sensor node from the base station is less than or equal to $d_{avg}$ then DE-LEACH uses the Equation (12) to find the threshold value T (n) for cluster head election.

$$T(n) = \begin{pmatrix} \frac{p_{opt1}}{1 - p \times \left(r \bmod \frac{1}{p}\right)} \times \left(c \times \frac{d_{avg}}{d_i}\right) & if\ n \in G \\ 0 & otherwise \end{pmatrix} \qquad (12)$$

Where $p_{opt1}$ is the desired percentage of nodes that can become cluster heads in a round and its value is 0.06250, r is the current round and c is a constant whose value is determined through simulation and $G$ is the set of nod that is not elected as a cluster head in the last $\frac{1}{p}$ rounds. The value of $p$ similar to LEACH protocol is 0.05.

Now when the node distance $d_i$ is greater than the average distance $d_{avg}$ of sensor nodes from base station then DE-LEACH uses the residual energy of nodes to elect cluster head (Equation 13) so that the nodes which have the sufficient energy will only respond for this duty. The initial energy of the nodes is used as reference energy.

$$T(n) = \begin{pmatrix} \frac{p_{opt2}}{1 - p \times \left(r \bmod \frac{1}{p}\right)} \times \frac{E_{i(r)}}{E_{i(t)}} & if\ n \in G \\ 0 & otherwise \end{pmatrix} \qquad (13)$$

Where $p_{opt2}$ is the desired percentage of nodes that can become cluster heads in a round and its value is 0.03125, r is the current round, $E_{i(r)}$ is the residual energy of a node, $E_{i(t)}$ is the node initial energy and $G$ is the set of node that is not elected as a cluster head in the last $\frac{1}{p}$ rounds. The value of $p$ similar to LEACH protocol is 0.05.

**Table 1: Radio Parameters used in DE-LEACH**

| Parameter | Value |
| --- | --- |
| $E_{elec}$ | 5 nJ/bit |
| $\varepsilon_{fs}$ | 10 pJ/bit/m$^2$ |
| $\varepsilon_{mp}$ | 0.0013 pJ/bit/m$^4$ |
| $E_0$ | 0.5 J |
| $E_{DA}$ | 5 nJ/bit/message |
| Message Size | 4000 bits |
| $p_{opt1}$ | 0.06250 |
| $p_{opt2}$ | 0.03125 |
| p | 0.05 |
| $d_0$ | 70m |

## 6. SIMULATION AND RESULTS

The performance of DE-LEACH is compared with LEACH. For simulation 100 x 100 square meters region with 100 sensor nodes are used as shown in Figure 1. In the Figure normal nodes are denoted by using the symbol (o) and the base station by (x). Radio parameters used for the simulation are given in Table 1. The following performance metrics are used for evaluating the protocol.

(i) **Network Lifetime:** This is the time interval between network operation start until the death of the last node.

(ii) **Stability Period:** This is the time interval between network operation start until the death of the first node.

(iii) **Number of Alive Nodes per round:** This will measure the number of live nodes in each round.

(iv) **Number of cluster heads per round:** This will measure the number of cluster heads formed in every round.

(v) **Number of packets sends to base station:** This will measure the total number of packets which are sent to base station.

The performance of DE-LEACH is evaluated by varying the value of constant c from 1 to 10 and the number of sensors from 10 to 100. Through simulation of 50 random topologies it is found that the optimal value of constant **c** for the system is 6. DE-LEACH equally distributes the energy load among the sensors, increases the stability period and throughput of the sensor network. Figure 2 shows that first and last node dies later in DE-LEACH as compared to LEACH. Thus network stability period and lifetime of DE-LEACH is more than LEACH. Figure 3 plots the number of live nodes per round and they are more in DE-LEACH as compared to LEACH. Figure 4 shows that total number of packets send to base station per round is more in DE-LEACH than LEACH. Figure 5 plots the total remaining energy per round of DE-LEACH and LEACH. From Figure 4 and 5 it is crystal clear that throughput and total remaining energy of DE-LEACH is more than LEACH.

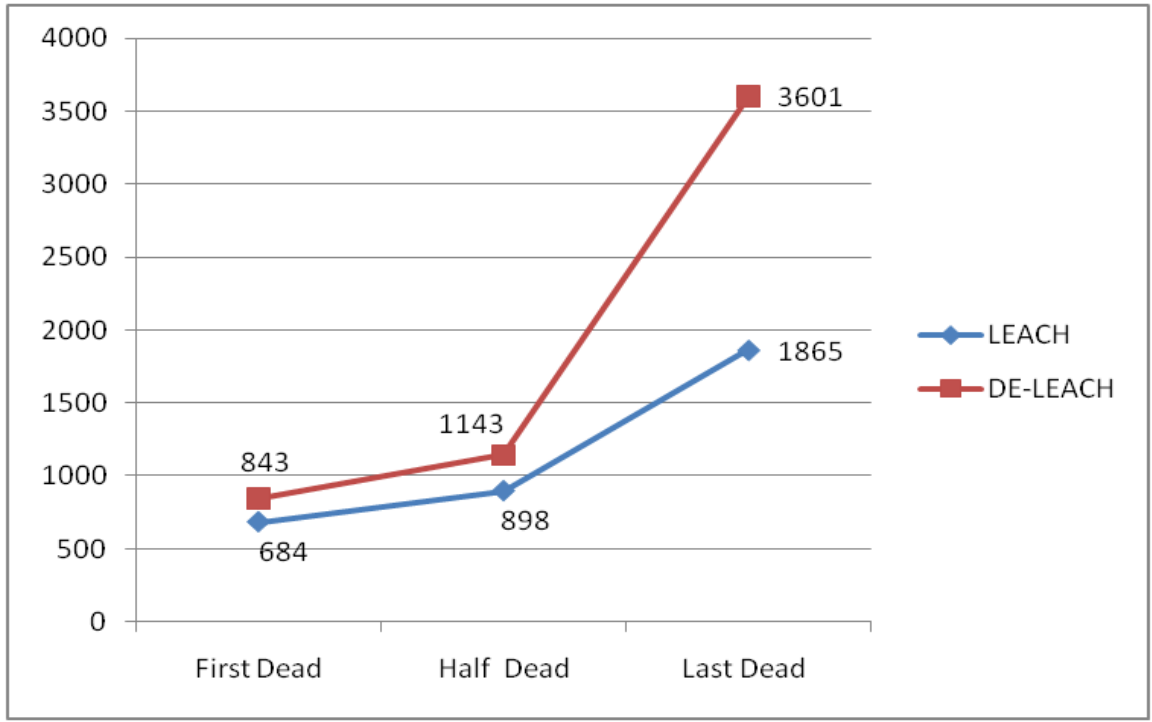

**Figure 2: Rounds for 1st, Half and Last Node Dead in DE-LEACH & LEACH**

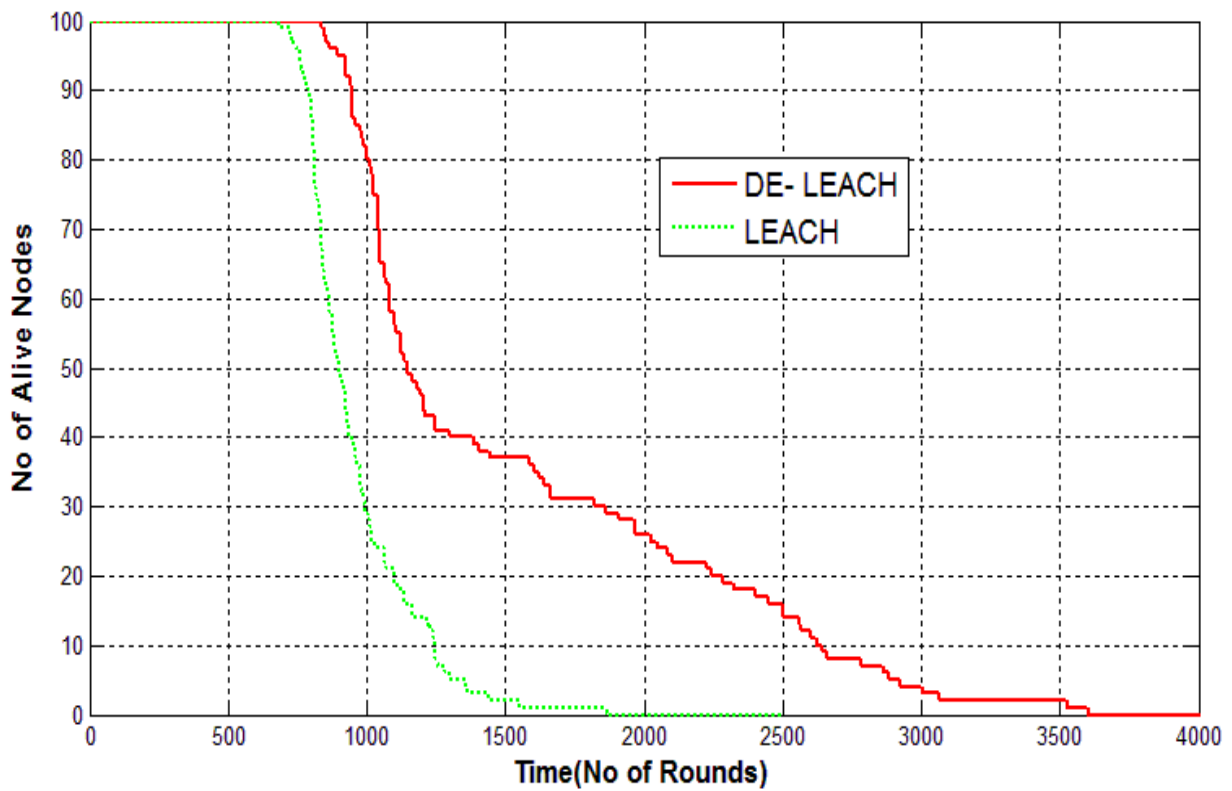

**Figure 3: Comparison of Number of Alive Nodes in DE-LEACH & LEACH**

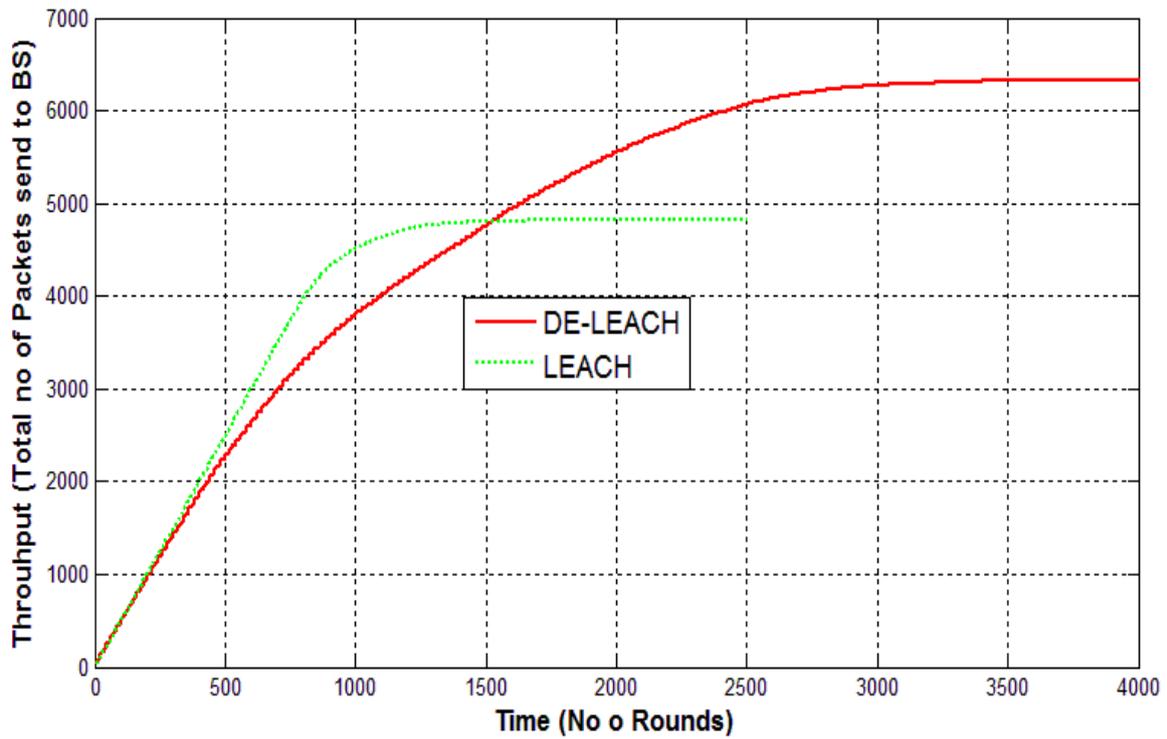

**Figure 4: Comparison of Throughput of DE-LEACH & LEACH**

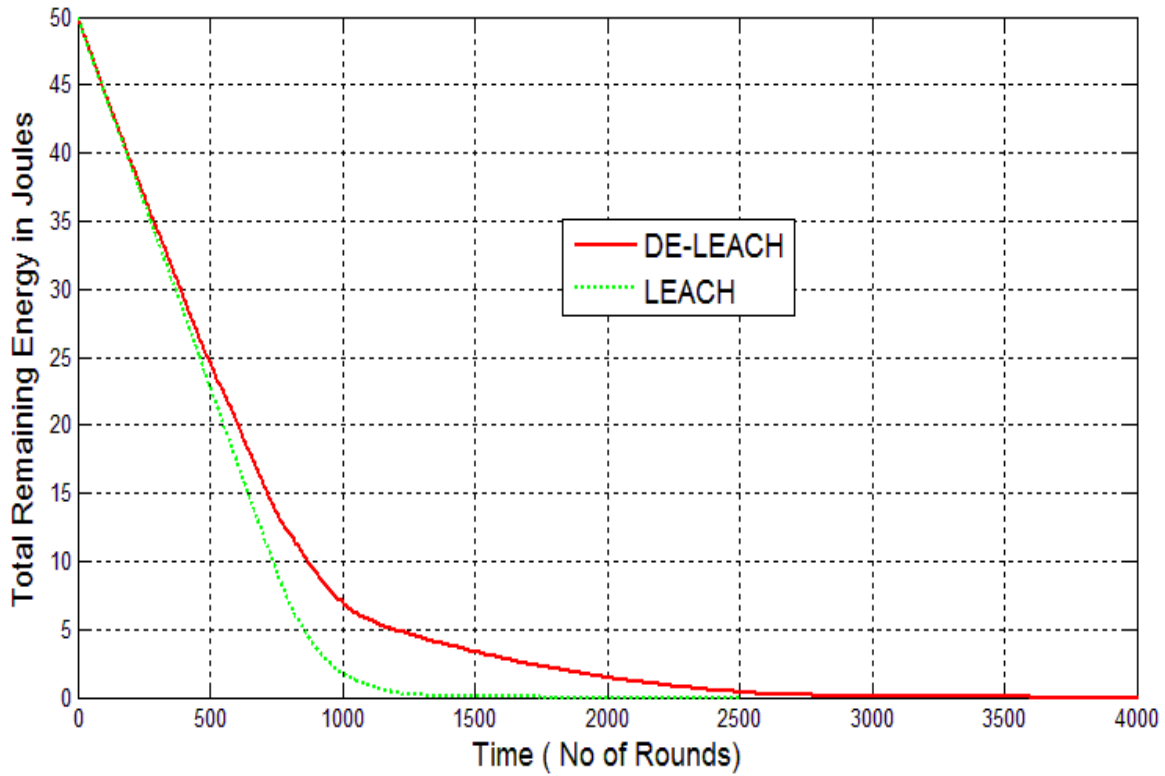

**Figure 5: Comparison of Total Remaining Energy Per Round in DE-LEACH & LEACH**

## 7. CONCLUSION
In this paper DE-LEACH a single hop communication protocol for homogeneous network is described. It improves the network lifetime, stable region and throughput of sensor network. For increasing the network energy efficiency it uses a residual energy and distance based cluster head election scheme. DE-LEACH ensures that nodes which are far away from base station will become cluster head only when they have sufficient energy for performing this duty and nearby nodes particularly in the mid of the sensing region have the highest probability to become a cluster head in a round. Simulation result shows that the proposed scheme is better than LEACH in energy efficiency and network life.